\documentclass[reprint,amsmath,amssymb,aps,superscriptaddress]{revtex4-2}

\usepackage{graphicx}
\usepackage{amsmath}
\usepackage{amssymb}
\usepackage{bm}
\usepackage{float}
\usepackage{tikz}
\makeatletter
\newcommand{\Needspace}[1]{%
  \par\begingroup
  \setlength{\dimen@}{#1}%
  \vskip\z@\@plus\dimen@
  \penalty-100\relax
  \vskip\z@\@plus-\dimen@
  \vskip\dimen@
  \penalty9999\relax
  \vskip-\dimen@
  \endgroup}
\newenvironment{fixedfigure}[1][0.2\textheight]
  {\Needspace{#1}\noindent\begin{minipage}{\columnwidth}%
   \def\@captype{figure}\centering}
  {\end{minipage}\par\medskip}
\makeatother

\begin{document}
\raggedbottom

\title{Complementary Approach to Anisotropic Flows in Heavy-Ion Collisions}

\author{E. Dlin}
 \affiliation{Moscow Institute of Physics and Technology, Dolgoprudny, Russia}
 \affiliation{A.~I.~Alikhanyan National Science Laboratory (AANL), Yerevan, Armenia}
\author{O. Teryaev}
 \affiliation{Moscow Institute of Physics and Technology, Dolgoprudny, Russia}
 \affiliation{Joint Institute for Nuclear Research, Dubna, Russia}
\date{June 10, 2026}

\begin{abstract}
We present a complementary count of directed flow \(v_1\) and elliptic flow \(v_2\). It does not replace known methods. A fixed laboratory plane defines up--down and left--right asymmetries. A scan of the test angle is written down as a theoretical proposal. Each event is given one angle \(\alpha\) drawn uniformly on \([0,2\pi)\), the fixed plane is rotated by \(\alpha\), and the asymmetries are averaged over events. With \(\theta=\alpha\) known, the linear fit recovers \(v_n^{\mathrm{true}}\). The quadratic mean, Eq.~(5), agrees with the pair form, Eq.~(21), in which the \(1/N\) self term cancels, and both sit above \(v_n^{\mathrm{true}}\) because the event noise in \(A\) is squared. The time for one count grows linearly with the number of particles. The pair sum, and the usual two-particle correlation, grow as the square of that number. The sample is PHSD Au+Au at \(\sqrt{s_{NN}}=9.2\,\mathrm{GeV}\), \(b=4.0\)--\(5.2\,\mathrm{fm}\).
\end{abstract}

\maketitle

\section{INTRODUCTION}

Anisotropic flows in heavy-ion collisions \cite{Poskanzer:1998yz} serve as a sensitive probe of early-stage dynamics and transport properties of the quark-gluon plasma produced in these extreme conditions. The coefficients \(v_{1}\) (directed flow) and \(v_{2}\) (elliptic flow) quantify the collective expansion of the created medium.

A particularly straightforward approach to extract anisotropic flow is to count particles that are emitted in specific directions relative to a fixed detector orientation. For example, one can define an up-down asymmetry, counting particles with \(\cos\phi > 0\) versus \(\cos\phi < 0\), and a left-right asymmetry, counting particles with \(\sin\phi > 0\) versus \(\sin\phi < 0\). These simple count asymmetries are head on related to the directed flow \(v_1\). Similarly, one can define asymmetries with respect to rotated planes to access elliptic flow \(v_2\).
 These quantities may be considered as the analogs of forward-backward and center-edge asymmetries \cite{Gounaris1993,Dvergsnes:2004tw} considered for azimuthal rather than polar angles.
 While these asymmetries provide event-averaged flow values when the detector plane is fixed, the same idea can be extended by scanning a test angle, as discussed below. The count is a complement to an event-plane analysis, not a substitute for one.

To test the method, we use the Parton-Hadron-String Dynamics (PHSD) model~\cite{PHSD, PHSDTwo} to simulate Au+Au collisions at \(\sqrt{s_{NN}} = 9.2\) GeV at freeze-out (\(b=4.0\)--\(5.2\) fm). The laboratory split is held fixed. The comparison is always the true flow coefficient \(v_n^{\mathrm{true}}=\langle\cos n\varphi\rangle\) from the generated momenta. While previous PHSD calculations~\cite{Prev} have demonstrated the rise of azimuthal anisotropies as a signature of quark-gluon plasma formation, here we concentrate on converting the count to that coefficient, and on relating the squared counting asymmetry to two-particle correlations.

\section{ASYMMETRY METHOD FOR DIRECTED AND ELLIPTIC FLOW}

\subsection{Fixed-plane asymmetries for directed flow}

Consider a detector that defines a fixed reference plane, which we align with the xz plane (i.e., the \(x\)-axis in the transverse plane). For each event, we count particles that are emitted in specific azimuthal directions relative to this plane. Define two simple count asymmetries:

\[
\begin{aligned}
A_{\mathrm{ud}} &= \frac{N_{\uparrow} - N_{\downarrow}}{N_{\uparrow} + N_{\downarrow}}, \quad
\begin{array}{l}
\uparrow: \cos\phi > 0, \\
\downarrow: \cos\phi < 0;
\end{array} \\[4pt]
A_{\mathrm{lr}} &= \frac{N_{\leftarrow} - N_{\rightarrow}}{N_{\leftarrow} + N_{\rightarrow}}, \quad
\begin{array}{l}
\leftarrow: \sin\phi > 0, \\
\rightarrow: \sin\phi < 0.
\end{array}
\end{aligned}
\tag{1}
\]

Figure~\ref{fig:fixed_v1} illustrates these definitions. The up–down asymmetry \(A_{\mathrm{ud}}\) counts particles that flew in the direction of the positive \(x\)-axis versus the negative \(x\)-axis (vertical split), while the left–right asymmetry \(A_{\mathrm{lr}}\) counts particles emitted with positive versus negative \(y\)-component (horizontal split).

\begin{fixedfigure}
\centering
\begin{tikzpicture}[scale=0.4]
\tikzset{region/.style={opacity=0.3}}

% A_lr
\begin{scope}[xshift=-2.5cm, yshift=0cm]
\draw[thick] (0,0) circle (2);
\draw[->] (-2.2,0) -- (2.2,0) node[right] {$x$};
\draw[->] (0,-2.2) -- (0,2.2) node[above] {$y$};
% Right half
\fill[blue!20, region] (0,0) -- (-90:2) arc (-90:90:2) -- cycle;
% Left half
\fill[red!20, region] (0,0) -- (90:2) arc (90:270:2) -- cycle;
% Labels
\node at (90:1.5) {$\leftarrow$};
\node at (270:1.5) {$\rightarrow$};
\node[blue] at (30:0.8) {$+$};
\node[red] at (210:0.8) {$-$};
\end{scope}

% A_ud
\begin{scope}[xshift=2.5cm, yshift=0cm]
\draw[thick] (0,0) circle (2);
\draw[->] (-2.2,0) -- (2.2,0) node[right] {$x$};
\draw[->] (0,-2.2) -- (0,2.2) node[above] {$y$};
% Upper half
\fill[green!20, region] (0,0) -- (0:2) arc (0:180:2) -- cycle;
% Lower half
\fill[orange!20, region] (0,0) -- (180:2) arc (180:360:2) -- cycle;
% Labels
\node at (0:1.5) {$\uparrow$};
\node at (180:1.5) {$\downarrow$};
\node[green] at (60:0.8) {$+$};
\node[orange] at (240:0.8) {$-$};
\end{scope}
\end{tikzpicture}
\caption{Fixed-plane asymmetries for directed flow: left: \(A_{\mathrm{ud}}\) (up/down) – right half (\(\cos\phi>0\), \(+\)) in blue, left half (\(\cos\phi<0\), \(-\)) in red; right: \(A_{\mathrm{lr}}\) (left/right) – upper half (\(\sin\phi>0\), \(+\)) in green, lower half (\(\sin\phi<0\), \(-\)) in orange. The detector plane is aligned with the \(x\)-axis.}
\label{fig:fixed_v1}
\end{fixedfigure}

How does this converge to flow coefficients? The azimuthal distribution of particles in a single event can be expanded in harmonics relative to the true reaction plane angle \(\Psi_{\mathrm{RP}}\):

\[
\frac{dN}{d\phi} \propto 1 + 2v_1 \cos(\phi - \Psi_{\mathrm{RP}}) + 2v_2 \cos 2(\phi - \Psi_{\mathrm{RP}}) + \cdots,
\tag{2}
\]

where \(v_1, v_2, \dots\) are the flow harmonics. The count \(A_{\mathrm{ud}}\) is the average of the square wave \(s=\mathrm{sign}(\cos\phi)\). Its Fourier series is \((4/\pi)\sum_{k=0}^{\infty}(-1)^k\cos((2k+1)\phi)/(2k+1)\). Directly substituting that series into Eq.~(2) gives
\[
\begin{aligned}
A_{\mathrm{ud}} &= \frac{4}{\pi} v_1 \cos\Psi_{\mathrm{RP}} - \frac{4}{3\pi} v_3 \cos 3\Psi_{\mathrm{RP}} \\
&\quad + \frac{4}{5\pi} v_5 \cos 5\Psi_{\mathrm{RP}} - \cdots, \\
A_{\mathrm{lr}} &= \frac{4}{\pi} v_1 \sin\Psi_{\mathrm{RP}} - \frac{4}{3\pi} v_3 \sin 3\Psi_{\mathrm{RP}} \\
&\quad + \frac{4}{5\pi} v_5 \sin 5\Psi_{\mathrm{RP}} - \cdots.
\end{aligned}
\tag{3}
\]
The same \(4/\pi\) is the coefficient of every leading term below. Dropping \(v_3,v_5,\ldots\) leaves
\[
A_{\mathrm{ud}} \approx \frac{4}{\pi} v_1 \cos\Psi_{\mathrm{RP}}, \qquad
A_{\mathrm{lr}} \approx \frac{4}{\pi} v_1 \sin\Psi_{\mathrm{RP}}.
\tag{4}
\]
From where one can easily obtain the flow coefficient straight away:

\[
v_{1}^{\mathrm{(est)}} = \frac{\pi}{4} \sqrt{A_{\mathrm{ud}}^2 + A_{\mathrm{lr}}^2}.
\tag{5}
\]

\subsection{Fixed-plane asymmetries for elliptic flow}

Elliptic flow can be accessed by defining asymmetries with respect to the same fixed plane and a plane rotated by \(45^\circ\). Let

\[
\begin{aligned}
A_{1} &= \frac{N_{\mathrm{in}} - N_{\mathrm{out}}}{N_{\mathrm{in}} + N_{\mathrm{out}}}, \quad
\begin{array}{l}
\mathrm{in:}\; |\phi| < \pi/4 \;\text{or}\; |\phi - \pi| < \pi/4, \\
\mathrm{out:}\; \text{otherwise};
\end{array} \\[4pt]
A_{2} &= \text{same as } A_1 \text{ but with the plane rotated by } \pi/4,
\end{aligned}
\tag{6}
\]

i.e., the “in” region for \(A_2\) is defined by \(|\phi - \pi/4| < \pi/4\) or \(|\phi - 5\pi/4| < \pi/4\). Figure~\ref{fig:fixed_v2} illustrates these definitions.

\begin{fixedfigure}
\centering
\begin{tikzpicture}[scale=0.4]
\tikzset{region/.style={opacity=0.3}}

% A_1
\begin{scope}[xshift=-2.5cm, yshift=0cm]
\draw[thick] (0,0) circle (2);
\draw[dashed] (0,0) -- (45:2.2);
\draw[dashed] (0,0) -- (-45:2.2);
\draw[dashed] (0,0) -- (135:2.2);
\draw[dashed] (0,0) -- (-135:2.2);
% In-plane regions
\fill[purple!20, region] (0,0) -- (-45:2) arc (-45:45:2) -- cycle;
\fill[purple!20, region] (0,0) -- (135:2) arc (135:225:2) -- cycle;
% Out-of-plane regions
\fill[yellow!20, region] (0,0) -- (45:2) arc (45:135:2) -- cycle;
\fill[yellow!20, region] (0,0) -- (-135:2) arc (-135:-45:2) -- cycle;
% Labels
\node at (0:0.9) {in};
\node at (180:0.9) {in};
\node at (90:1.3) {out};
\node at (-90:1.3) {out};
\end{scope}

% A_2
\begin{scope}[xshift=2.5cm, yshift=0cm]
\draw[thick] (0,0) circle (2);
\draw[->] (-2.2,0) -- (2.2,0) node[right] {$x$};
\draw[->] (0,-2.2) -- (0,2.2) node[above] {$y$};
% In-plane regions
\fill[green!20, region] (0,0) -- (0:2) arc (0:90:2) -- cycle;
\fill[green!20, region] (0,0) -- (180:2) arc (180:270:2) -- cycle;
% Out-of-plane regions
\fill[orange!20, region] (0,0) -- (90:2) arc (90:180:2) -- cycle;
\fill[orange!20, region] (0,0) -- (-90:2) arc (-90:0:2) -- cycle;
% Labels
\node at (45-15:0.9) {in};
\node at (225-15:0.9) {in};
\node at (135-10:1.3) {out};
\node at (-45+10:1.3) {out};
\end{scope}
\end{tikzpicture}
\caption{Fixed-plane asymmetries for elliptic flow: left: \(A_1\) (in/out relative to the \(x\)-axis) – purple (in) and yellow (out); right: \(A_2\) (in/out relative to the plane rotated by \(45^\circ\)) – green (in) and orange (out).}
\label{fig:fixed_v2}
\end{fixedfigure}

A similar derivation gives
\[
A_1 \approx \frac{4}{\pi} v_2 \cos 2\Psi_{\mathrm{RP}}, \qquad
A_2 \approx \frac{4}{\pi} v_2 \sin 2\Psi_{\mathrm{RP}},
\tag{7}
\]
and once again:

\[
v_{2}^{\mathrm{(est)}} = \frac{\pi}{4} \sqrt{A_1^2 + A_2^2}.
\tag{8}
\]

\subsection{Higher harmonics}

Equation~(3) keeps the higher odd harmonics. At one angle the two directed asymmetries add to
\[
A_{\mathrm{ud}}^2 + A_{\mathrm{lr}}^2 = \left(\frac{4}{\pi}\right)^2 \Big[
v_1^2 + \frac{1}{9}v_3^2 + \frac{1}{25}v_5^2 + \cdots + \text{cross terms}
\Big].
\tag{10}
\]
A cross term is a product of \(\cos m\Psi_{\mathrm{RP}}\) and \(\cos n\Psi_{\mathrm{RP}}\), or of the corresponding sines, with \(m\neq n\). When \(\Psi_{\mathrm{RP}}\) is spread uniformly over \([0,2\pi)\), those products average to zero, and \(\langle\cos^2 n\Psi_{\mathrm{RP}}\rangle=\langle\sin^2 n\Psi_{\mathrm{RP}}\rangle=1/2\). Each asymmetry then averages to
\[
\langle A_{\mathrm{ud}}^2 \rangle = \langle A_{\mathrm{lr}}^2 \rangle
= \frac{8}{\pi^2} \left( v_1^2 + \frac{1}{9}v_3^2 + \frac{1}{25}v_5^2 + \cdots \right).
\tag{11}
\]
The in and out regions of Eq.~(6) follow \(\cos 2\phi\). The next term of that series is \(\cos 6\phi\), so
\[
\begin{aligned}
A_{1} &= \frac{4}{\pi} v_2 \cos 2\Psi_{\mathrm{RP}} - \frac{4}{3\pi} v_6 \cos 6\Psi_{\mathrm{RP}} + \cdots, \\
A_{2} &= \frac{4}{\pi} v_2 \sin 2\Psi_{\mathrm{RP}} - \frac{4}{3\pi} v_6 \sin 6\Psi_{\mathrm{RP}} + \cdots,
\end{aligned}
\tag{13}
\]
and
\[
\langle A_{1}^2 \rangle = \langle A_{2}^2 \rangle
= \frac{8}{\pi^2} \left( v_2^2 + \frac{1}{9}v_6^2 + \cdots \right).
\tag{14}
\]
At one angle the leading terms are \(A_{\mathrm{ud}}^2+A_{\mathrm{lr}}^2=(16/\pi^2)v_1^2\) and \(A_1^2+A_2^2=(16/\pi^2)v_2^2\). This is why Eqs.~(5) and~(8) carry the factor \(\pi/4\).

If \(A\) is averaged over events before it is squared, the average vanishes. \(A\) follows \(v\cos\Psi_{\mathrm{RP}}\) or \(v\sin\Psi_{\mathrm{RP}}\), and the mean of a cosine or a sine over a uniform angle is zero. In the PHSD sample the impact parameter lies along \(x\), so \(\Psi_{\mathrm{RP}}=0\) in every event. The detector planes stay fixed, and each event draws an angle \(\alpha\) uniformly from \([0,2\pi)\) that rotates those planes. The angle between the detector and the reaction plane is \(\theta=\alpha\). On that event
\[
A_{\mathrm{ud}}=v\cos\theta,\qquad A_{\mathrm{lr}}=-v\sin\theta,
\]
and
\[
A_{1}=v\cos n\theta,\qquad A_{2}=-v\sin n\theta.
\]
The cosine form has no sine in it, and the sine form has no cosine in it. Since \(A=(4/\pi)v_n\) times the cosine or the sine, the fitted \(v\) is \((4/\pi)v_n\), and
\[
v_{n}^{(\mathrm{fit})}=\frac{\pi}{4}v
\tag{22}
\]
returns the same \(v_n\) as Eqs.~(4) and~(7). Equations~(5) and~(8) instead take \(\pi/4\) times the square root of the sum of the two mean squares. A measured asymmetry is the cosine term plus a remainder from the finite number of particles in the event. The product of the cosine and the remainder averages to zero. The square of the remainder remains, so Eqs.~(5) and~(8) lie above Eq.~(22).

\subsection{Scanning the test angle}

The same uniform angle can be built inside one event by moving a test angle \(\psi\) from \(0\) to \(2\pi\). The flow coefficients are then
\[
\begin{aligned}
v_{1}^{\mathrm{(noRP)}} &= \frac{\pi}{4} \sqrt{ \langle A_{\mathrm{ud}}^2(\psi) \rangle_{\psi} + \langle A_{\mathrm{lr}}^2(\psi) \rangle_{\psi} }, \\
v_{2}^{\mathrm{(noRP)}} &= \frac{\pi}{4} \sqrt{ \langle A_{1}^2(\psi) \rangle_{\psi} + \langle A_{2}^2(\psi) \rangle_{\psi} }.
\end{aligned}
\tag{15}
\]
The average over \(\psi\) removes the cross terms in the same way as Eqs.~(11) and~(14). Figure~\ref{fig:norp_regions} shows the four counts at one value of \(\psi\). Figure~\ref{fig:event_scatter} shows that the two asymmetries of one harmonic contribute equally: each ratio sits near \(1/2\). The PHSD numbers in Sec.~\ref{sec:results} are the mean over events. Each event carries its own angle \(\alpha\), as set in Sec.~\ref{sec:direct}.

\begin{fixedfigure}
\centering
\begin{tikzpicture}[scale=0.4]
% Common style
\tikzset{
    testplane/.style={->, thick, blue},
    region/.style={opacity=0.3},
}

% A_lr
\begin{scope}[xshift=-5cm, yshift=2.5cm]
\draw[thick] (0,0) circle (2);
\def\ang{30}
\draw[testplane] (0,0) -- (\ang:2.2) node[above right] {$\psi$};
\draw[dashed] (0,0) -- (\ang+180:2.2);
\draw[dashed] (0,0) -- (\ang+90:2.2);
\draw[dashed] (0,0) -- (\ang-90:2.2);
% cos>0
\fill[blue!20, region] (0,0) -- (\ang-90:2) arc (\ang-90:\ang+90:2) -- cycle;
% cos<0
\fill[red!20, region] (0,0) -- (\ang+90:2) arc (\ang+90:\ang+270:2) -- cycle;
% Labels
\node at (\ang+90:1.5) {$\leftarrow$};
\node at (\ang-90:1.5) {$\rightarrow$};
\node[blue] at (\ang+30:0.8) {$+$};
\node[red] at (\ang+210:0.8) {$-$};
\end{scope}

% A_ud
\begin{scope}[xshift=0cm, yshift=2.5cm]
\draw[thick] (0,0) circle (2);
\def\ang{30}
\draw[testplane] (0,0) -- (\ang:2.2) node[above right] {$\psi$};
\draw[dashed] (0,0) -- (\ang+180:2.2);
\draw[dashed] (0,0) -- (\ang+90:2.2);
\draw[dashed] (0,0) -- (\ang-90:2.2);
% sin>0
\fill[green!20, region] (0,0) -- (\ang:2) arc (\ang:\ang+180:2) -- cycle;
% sin<0
\fill[orange!20, region] (0,0) -- (\ang+180:2) arc (\ang+180:\ang+360:2) -- cycle;
% Labels
\node at (\ang:1.5) {$\uparrow$};
\node at (\ang+180:1.5) {$\downarrow$};
\node[green] at (\ang+60:0.8) {$+$};
\node[orange] at (\ang+240:0.8) {$-$};
\end{scope}

% A_1
\begin{scope}[xshift=-5cm, yshift=-3cm]
\draw[thick] (0,0) circle (2);
\draw[->] (-2.2,0) -- (2.2,0) node[right] {$x$};
\draw[->] (0,-2.2) -- (0,2.2) node[above] {$y$};
\def\ang{30}
\draw[testplane] (0,0) -- (\ang:2.2) node[above right] {$\psi$};
\draw[dashed] (0,0) -- (\ang+45:2.2);
\draw[dashed] (0,0) -- (\ang-45:2.2);
\draw[dashed] (0,0) -- (\ang+135:2.2);
\draw[dashed] (0,0) -- (\ang-135:2.2);
% In-plane regions
\fill[purple!20, region] (0,0) -- (\ang-45:2) arc (\ang-45:\ang+45:2) -- cycle;
\fill[purple!20, region] (0,0) -- (\ang+135:2) arc (\ang+135:\ang+225:2) -- cycle;
% Out-of-plane regions
\fill[yellow!20, region] (0,0) -- (\ang+45:2) arc (\ang+45:\ang+135:2) -- cycle;
\fill[yellow!20, region] (0,0) -- (\ang-135:2) arc (\ang-135:\ang-45:2) -- cycle;
% Labels
\node at (\ang-15:0.9) {in};
\node at (\ang+180-15:0.9) {in};
\node at (\ang+90+10:1.3) {out};
\node at (\ang-90-10:1.3) {out};
\end{scope}

% A_2
\begin{scope}[xshift=0cm, yshift=-3cm]
\draw[thick] (0,0) circle (2);
\draw[->] (-2.2,0) -- (2.2,0) node[right] {$x$};
\draw[->] (0,-2.2) -- (0,2.2) node[above] {$y$};
\def\ang{30}
\draw[testplane] (0,0) -- (\ang:2.2) node[above right] {$\psi$};
% In-plane regions
\fill[green!20, region] (0,0) -- (\ang:2) arc (\ang:\ang+90:2) -- cycle;
\fill[green!20, region] (0,0) -- (\ang+180:2) arc (\ang+180:\ang+270:2) -- cycle;
% Out-of-plane regions
\fill[orange!20, region] (0,0) -- (\ang+90:2) arc (\ang+90:\ang+180:2) -- cycle;
\fill[orange!20, region] (0,0) -- (\ang-90:2) arc (\ang-90:\ang:2) -- cycle;
% Labels
\node at (\ang+45-15:0.9) {in};
\node at (\ang-135+15:0.9) {in};
\node at (\ang-45+10:1.3) {out};
\node at (\ang+135-10:1.3) {out};
\end{scope}
\end{tikzpicture}
\caption{The four asymmetries at one test angle \(\psi\) (blue arrow). Top left: \(A_{\mathrm{lr}}\), split perpendicular to \(\psi\). Top right: \(A_{\mathrm{ud}}\), split along \(\psi\). Bottom left: \(A_1\), in and out relative to \(\psi\). Bottom right: \(A_2\), in and out relative to \(\psi+\pi/4\).}
\label{fig:norp_regions}
\end{fixedfigure}

\begin{fixedfigure}[0.62\textheight]
\centering
\includegraphics[width=\columnwidth]{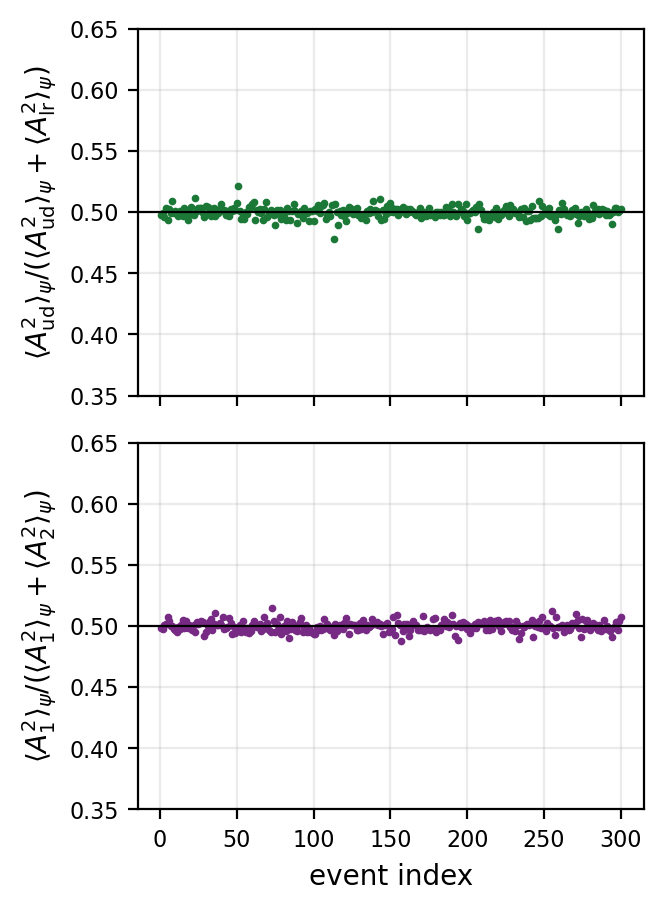}
\caption{Scan of the test angle in Eq.~(15), for 300 events. Top: \(\langle A_{\mathrm{ud}}^2 \rangle_{\psi} / (\langle A_{\mathrm{ud}}^2 \rangle_{\psi} + \langle A_{\mathrm{lr}}^2 \rangle_{\psi})\). Bottom: the same ratio for \(A_1\) and \(A_2\). Both ratios sit near \(1/2\).}
\label{fig:event_scatter}
\end{fixedfigure}

\section{COMPARISON WITH THE MODEL REACTION PLANE}\label{sec:direct}

The comparison is the true flow coefficient from the generated momenta, with the same cuts as the count,
\[
v_{n}^{\mathrm{true}}
=\langle \cos n\varphi\rangle,
\qquad
\varphi=\mathrm{atan2}(p_y,p_x).
\tag{16}
\]
In PHSD the impact parameter lies along \(x\), so the reaction plane is \(\Psi_{\mathrm{RP}}=0\) in every event. The detector planes stay fixed. Each event draws one angle \(\alpha\) uniformly from \([0,2\pi)\), and the planes are rotated by \(\alpha\). Relative to the detector, the reaction plane is then uniform over the sample. Equation~(16) is evaluated on the generated azimuths before this rotation. In Fig.~\ref{fig:A_theta} the angle of each event is \(\theta=\alpha\).

The generated list covers the full solid angle. The particles kept for the count have \(p_T>0.2\,\mathrm{GeV}/c\) and \(|\eta|<2.27\), on both sides of midrapidity, out to the beam rapidity at \(9.2\,\mathrm{GeV}\). Directed flow changes sign between the two hemispheres. Each particle enters \(v_1\) and the directed counts with weight \(\mathrm{sign}(\eta)\), so the two sides add~\cite{Adamczyk:2014ipa}. Elliptic flow has the same sign on both sides and is counted with weight one. On this sample \(|v_1^{\mathrm{true}}|=0.01992\) and \(v_2^{\mathrm{true}}=0.02431\). The same cuts and the same weight enter Eqs.~(5), (8), (21), and~(22).

\section{RESULTS AND DISCUSSION}\label{sec:results}

Each event is given one angle \(\alpha\), as in Sec.~\ref{sec:direct}. The asymmetries at that angle are averaged as the number of events grows.

Figure~\ref{fig:A_theta} shows every event at \(\theta=\alpha\). The fits are \(A_{\mathrm{ud}}=v\cos\theta\), \(A_{\mathrm{lr}}=-v\sin\theta\), \(A_1=v\cos 2\theta\), and \(A_2=-v\sin 2\theta\). The flow coefficient is \((\pi/4)v\), Eq.~(22). For directed flow the two fits give \(0.02000\) and \(0.01951\), and \(|v_1^{\mathrm{true}}|=0.01992\). For elliptic flow they give \(0.02446\) and \(0.02428\), and \(v_2^{\mathrm{true}}=0.02431\).

Figure~\ref{fig:eq5_eq21} shows Eqs.~(5) and~(21) for directed flow as \(N_{\mathrm{ev}}\) grows. Equation~(5) squares the two directed asymmetries at the angle \(\alpha\). Equation~(21) builds the same square from the angles between particle pairs. The two curves lie on top of each other. Both sit above the fits of Fig.~\ref{fig:A_theta}, because the square of the event-by-event remainder remains after the \(1/N\) terms cancel in Eq.~(21).

\begin{fixedfigure}[0.42\textheight]
\centering
\includegraphics[width=\columnwidth]{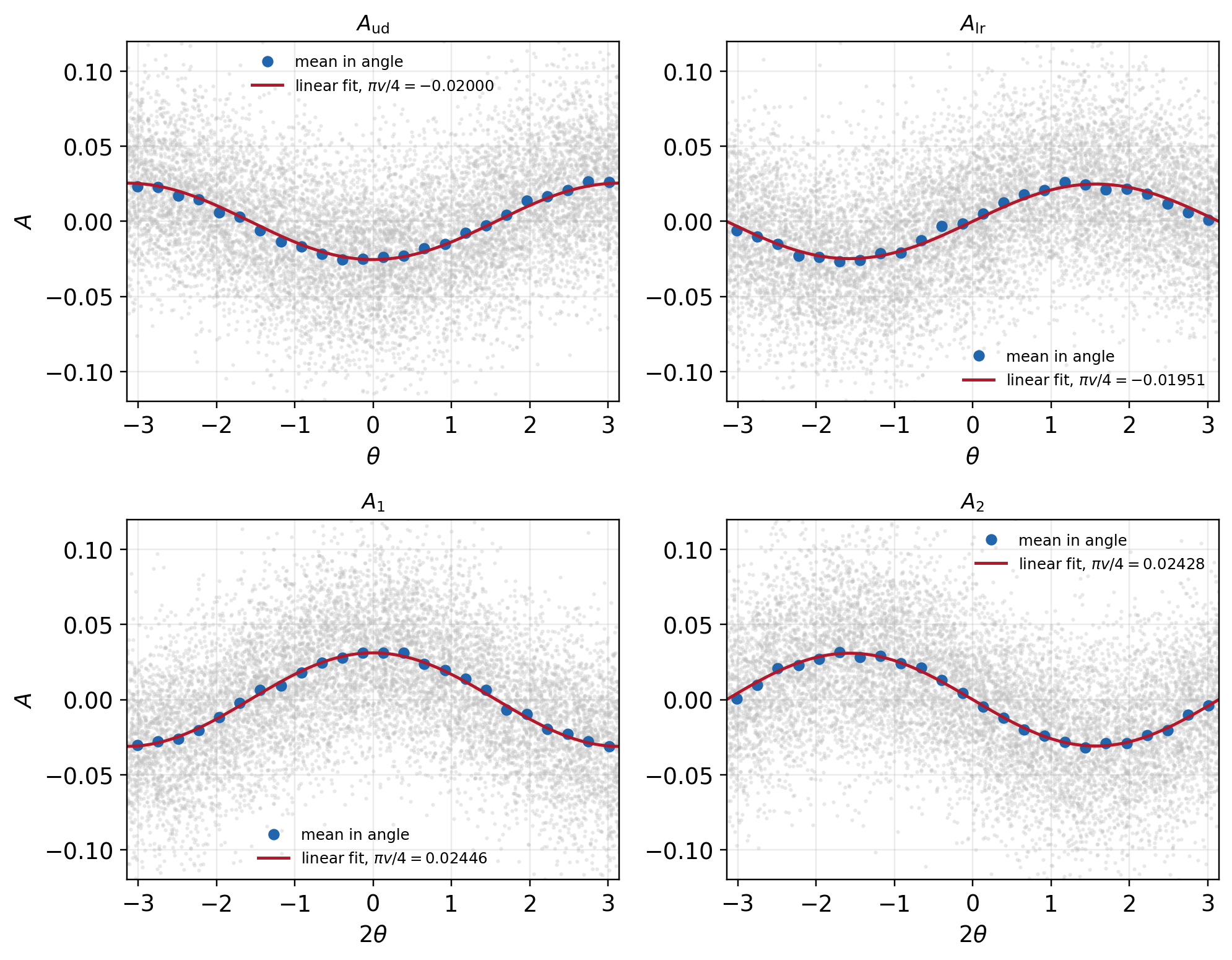}
\caption{Asymmetry of each event at the angle \(\theta=\alpha\) between the detector plane and the reaction plane. Points are single events. Blue: mean in the angle bin. Red: the fit \(A_{\mathrm{ud}}=v\cos\theta\), \(A_{\mathrm{lr}}=-v\sin\theta\), \(A_1=v\cos 2\theta\), \(A_2=-v\sin 2\theta\), shown as \((\pi/4)v\). Top: directed flow, \(0.02000\) and \(0.01951\), with \(|v_1^{\mathrm{true}}|=0.01992\). Bottom: elliptic flow, \(0.02446\) and \(0.02428\), with \(v_2^{\mathrm{true}}=0.02431\).}
\label{fig:A_theta}
\end{fixedfigure}

\begin{fixedfigure}[0.38\textheight]
\centering
\includegraphics[width=\columnwidth]{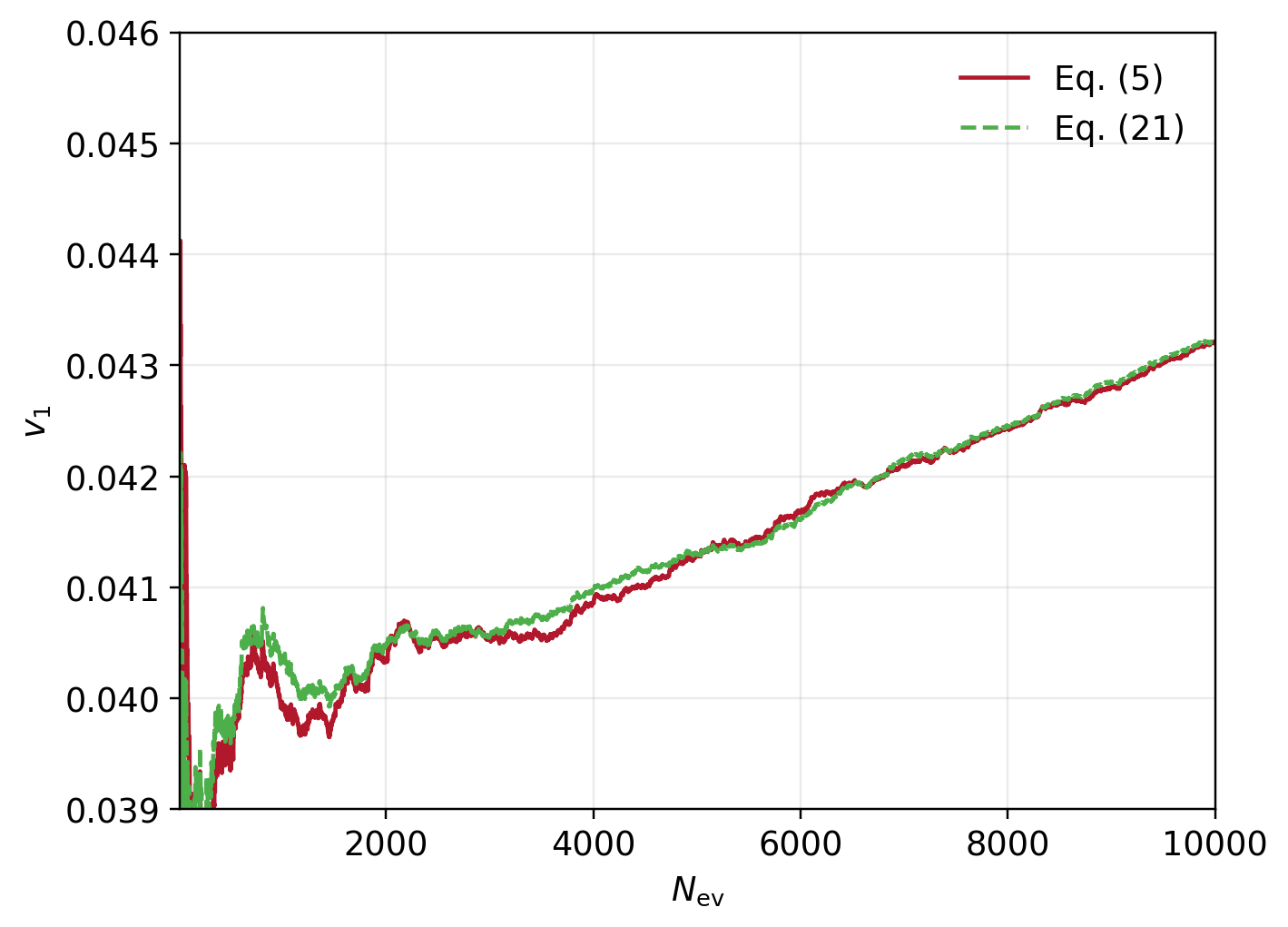}
\caption{Directed flow versus \(N_{\mathrm{ev}}\). Red: Eq.~(5). Green, dashed: Eq.~(21), \(\frac{\pi}{4}\sqrt{2\langle Q\rangle}\). The two curves agree. Cuts: \(p_T>0.2\,\mathrm{GeV}/c\), \(|\eta|<2.27\), weight \(\mathrm{sign}(\eta)\).}
\label{fig:eq5_eq21}
\end{fixedfigure}

\section{RELATION TO TWO-PARTICLE CORRELATIONS}\label{sec:twopc}

The up--down count is a sum of signs,
\[
A_{\mathrm{ud}}(\psi)=\frac{1}{N}\sum_{k=1}^{N}s_k(\psi),
\]
\[
s_k=\mathrm{sign}\bigl[\cos(\varphi_k-\psi)\bigr]=\pm1.
\tag{17}
\]
Then
\[
A_{\mathrm{ud}}=\frac{1}{N}\sum_{k=1}^{N}s_k,
\tag{18}
\]
and
\[
\begin{aligned}
A_{\mathrm{ud}}^2
&=\frac{1}{N^2}\sum_{k=1}^{N}\sum_{\ell=1}^{N}s_k s_\ell
=\frac{1}{N}+\frac{1}{N^2}\sum_{k\neq\ell}s_k s_\ell,
\end{aligned}
\tag{19}
\]
since \(s_k^2=1\). The term \(1/N\) comes from pairing each particle with itself. The second term sums pairs of different particles.

For two particles, \(s_i s_j=+1\) when the test plane does not lie between them, and \(s_i s_j=-1\) when it does. Figure~\ref{fig:pair_angle} shows the two cases. The shorter arc between the particles is \(\Delta=\min(\phi,2\pi-\phi)\). Averaging over the test angle gives
\[
\langle s_i s_j\rangle_{\psi}=1-\frac{2\Delta}{\pi}.
\tag{20}
\]
There are \(N(N-1)\) ordered pairs \(i\neq j\). Substituting Eq.~(20) into Eq.~(19) gives
\[
\begin{aligned}
\langle A_{\mathrm{ud}}^2\rangle_{\psi}
&=\frac{1}{N}+\frac{1}{N^2}\sum_{i\neq j}\left(1-\frac{2\Delta_{ij}}{\pi}\right)\\
&=\frac{1}{N}+\frac{N(N-1)}{N^2}
-\frac{2}{N^2}\sum_{i\neq j}\frac{\Delta_{ij}}{\pi}\\
&=\frac{1}{N}+\left(1-\frac{1}{N}\right)
-\frac{2}{N^2}\sum_{i\neq j}\frac{\Delta_{ij}}{\pi}\\
&=1-\frac{2}{N^2}\sum_{i\neq j}\frac{\Delta_{ij}}{\pi}.
\end{aligned}
\tag{21}
\]
The \(+1/N\) from a particle paired with itself and the \(-1/N\) inside \((N-1)/N\) cancel. Equation~(21) is what remains. The opening \(\Delta\) is at most \(\pi\), and there are fewer pairs than \(N^2\), so Eq.~(21) stays above \(-1\). If the azimuth is flat, the mean opening is \(\pi/2\). Each pair then averages to zero, and \(\langle A_{\mathrm{ud}}^2\rangle_{\psi}\) falls to \(1/N\), the size of the counting scatter in one event.

The same pair sum, taken for the up--down count and the left--right count together, is Eq.~(5). Figure~\ref{fig:eq5_eq21} shows the two curves on top of each other. The fit of Eq.~(22) follows \(v_1^{\mathrm{true}}\). Equations~(5) and~(21) lie above it: they square the scatter of each event, and that square does not average away.

The time is set by the number of particles in the event. One pass over those particles, the four counts on a fixed plane, grows linearly with \(N\). Equation~(21) visits every pair, so its time grows as \(N^2\). The two-particle correlation of Borghini, Dinh and Ollitrault~\cite{Borghini:2001vi,Borghini:2001xq} grows in the same way. Figure~\ref{fig:norp_scaling} puts the three on the same events, each cut to a fixed multiplicity. At large \(N\) the pair sums take far longer than the single count.

\begin{fixedfigure}
\centering
\begin{tikzpicture}[scale=0.55]
\tikzset{region/.style={opacity=0.25}}
% left: plane not between the two particles
\begin{scope}[xshift=-2.6cm]
\draw[thick] (0,0) circle (2);
\draw[dashed] (0,0) -- (90:2.2);
\draw[dashed] (0,0) -- (270:2.2);
\fill[green!25, region] (0,0) -- (0:2) arc (0:180:2) -- cycle;
\fill[orange!25, region] (0,0) -- (180:2) arc (180:360:2) -- cycle;
\draw[->, thick, blue] (0,0) -- (0:2.3) node[right] {$\psi$};
\filldraw[black] (40:1.65) circle (2.2pt) node[above right] {$i$};
\filldraw[black] (70:1.65) circle (2.2pt) node[above] {$j$};
\draw[thick] (40:1.15) arc (40:70:1.15);
\node at (55:0.75) {$+\phi$};
\node[green!50!black] at (20:0.7) {$+$};
\node[green!50!black] at (100:0.7) {$+$};
\node[font=\small] at (0,-2.55) {plane not between $i,j$};
\end{scope}
% right: plane between them
\begin{scope}[xshift=2.6cm]
\draw[thick] (0,0) circle (2);
\draw[dashed] (0,0) -- (90:2.2);
\draw[dashed] (0,0) -- (270:2.2);
\fill[green!25, region] (0,0) -- (0:2) arc (0:180:2) -- cycle;
\fill[orange!25, region] (0,0) -- (180:2) arc (180:360:2) -- cycle;
\draw[->, thick, blue] (0,0) -- (0:2.3) node[right] {$\psi$};
\filldraw[black] (50:1.65) circle (2.2pt) node[above right] {$i$};
\filldraw[black] (230:1.65) circle (2.2pt) node[below left] {$j$};
\draw[thick] (50:1.15) arc (50:230:1.15);
\node at (140:0.85) {$-(2\pi-\phi)$};
\node[green!50!black] at (40:0.7) {$+$};
\node[orange!70!black] at (220:0.7) {$-$};
\node[font=\small] at (0,-2.55) {plane between $i,j$};
\end{scope}
\end{tikzpicture}
\caption{How the pair angle enters \(s_i s_j\). The dashed line is the split at the test angle \(\psi\). Left: the plane does not separate the two particles, the arc is \(+\phi\), and \(s_i s_j=+1\). Right: the plane lies between them, the long arc is taken with a minus sign, and \(s_i s_j=-1\).}
\label{fig:pair_angle}
\end{fixedfigure}

\begin{fixedfigure}[0.52\textheight]
\centering
\includegraphics[width=\columnwidth]{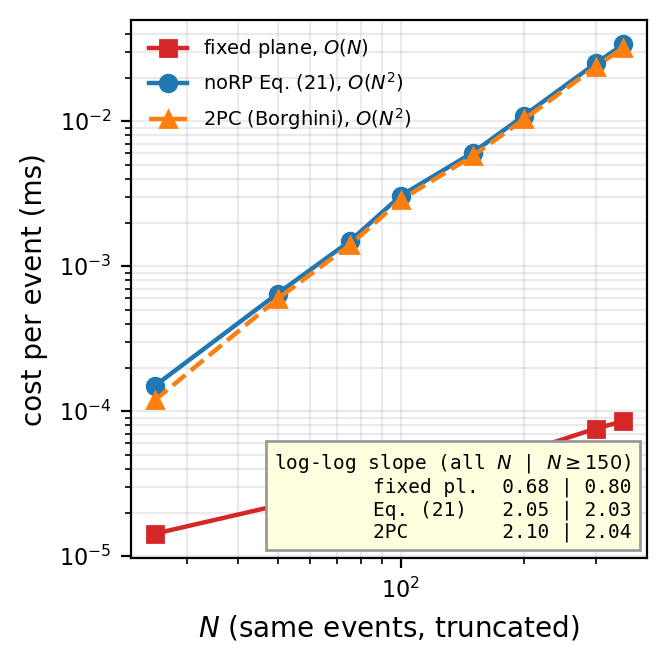}
\caption{Time per event versus multiplicity, on the same events cut to a fixed \(N\). One fixed plane, four counts: grows as \(N\). Eq.~(21), pair sum: grows as \(N^2\). Two-particle correlation~\cite{Borghini:2001vi,Borghini:2001xq}: grows as \(N^2\).}
\label{fig:norp_scaling}
\end{fixedfigure}

\section{CONCLUSIONS}

The asymmetries are a complement to the usual event-plane and pair analyses. Particles are sorted by which side of a fixed plane they fall on.

In the PHSD sample the reaction plane lies along \(x\). Each event rotates the detector plane by one random angle \(\alpha\) (Sec.~\ref{sec:direct}). With that angle known, the fit in Fig.~\ref{fig:A_theta} returns \(v_n^{\mathrm{true}}\). The mean of the squared asymmetries, Eqs.~(5) and~(8), lies above \(v_n^{\mathrm{true}}\), because the remainder in each event is squared. Equation~(21) is that same mean written as a sum over pairs: the \(1/N\) terms cancel, as shown between Eqs.~(19) and~(21). The pair sum and the standard two-particle correlation both grow as \(N^2\) (Fig.~\ref{fig:norp_scaling}). The scan over a test angle, Eq.~(15), is the same relation written inside one event. The numbers above are the mean over events.

\section*{ACKNOWLEDGMENTS}

We thank A.~Borissov, E.~Bratkovskaya, V. Riabov, I.~Serenkova, A. Taranenko and V.~Voronyuk for useful discussions and correspondence. Work is supported by RSF grant 25-72-30005


\begin{thebibliography}{99}
%\cite{Poskanzer:1998yz}
\bibitem{Poskanzer:1998yz}
A.~M.~Poskanzer and S.~A.~Voloshin,
%``Methods for analyzing anisotropic flow in relativistic nuclear collisions,''
Phys. Rev. C \textbf{58}, 1671-1678 (1998)
%doi:10.1103/PhysRevC.58.1671
[arXiv:nucl-ex/9805001 [nucl-ex]].
%1734 citations counted in INSPIRE as of 10 Apr 2026
\bibitem{Gounaris1993} G. Gounaris, J. Layssac, G. Moultaka, and F. M. Renard, 
%``Analytic expressions of cross sections, asymmetries and \(W\) density matrices for \(e^+e^- \to W^+W^-\) with general 3-boson couplings,'' 
Int. J. Mod. Phys. A \textbf{8}, 3285 (1993).

%\cite{Dvergsnes:2004tw}
\bibitem{Dvergsnes:2004tw}
E.~W.~Dvergsnes, P.~Osland, A.~A.~Pankov and N.~Paver,
%``Center edge asymmetry at hadron colliders,''
Phys. Rev. D \textbf{69}, 115001 (2004)
%doi:10.1103/PhysRevD.69.115001
[arXiv:hep-ph/0401199 [hep-ph]].
%40 citations counted in INSPIRE as of 10 Apr 2026


\bibitem{PHSD} W. Cassing and E. L. Bratkovskaya, Phys. Rev. C \textbf{78}, 034919 (2008).
\bibitem{PHSDTwo} W. Cassing and E.L. Bratkovskaya, Nucl. Phys. A \textbf{831}  215-242 (2009)
\bibitem{Prev} V. P. Konchakovski, E. L. Bratkovskaya, W. Cassing, V. D. Toneev, and V. Voronyuk, Phys. Rev. C \textbf{85}, 044922 (2012).
\bibitem{Kolb} Peter F. Kolb, Phys.Rev.C \textbf{68}, 031902 (2003)
\bibitem{Borghini:2001vi}
N.~Borghini, P.~M.~Dinh, and J.-Y.~Ollitrault,
Phys. Rev. C \textbf{63}, 054906 (2001)
[arXiv:nucl-th/0007063].
\bibitem{Borghini:2001xq}
N.~Borghini, P.~M.~Dinh, and J.-Y.~Ollitrault,
Phys. Rev. C \textbf{64}, 054901 (2001)
[arXiv:nucl-th/0105040].
\bibitem{Adamczyk:2014ipa}
L.~Adamczyk {\it et al.} (STAR Collaboration),
Phys. Rev. Lett. \textbf{112}, 162301 (2014)
[arXiv:1401.3043].
\end{thebibliography}
\end{document}